\documentclass[aps,prb,superscriptaddress,floatfix,doublecol,showpacs]{epl2}

\usepackage{verbatim}

\usepackage{amsmath}
\usepackage{amsfonts}
\usepackage{amssymb}
\usepackage{graphicx}
\usepackage{bm}
\usepackage{color}
\usepackage{multirow}

\newcommand{\be}{\begin{equation}}
\newcommand{\ee}{\end{equation}}
\newcommand{\bea}{\begin{eqnarray}}
\newcommand{\eea}{\end{eqnarray}}
\newcommand{\ColorOnline}{(Color on-line) }

\newcommand{\HH}{\hat H}

\newcommand{\hNd}{{\hat N}_\text{d}}

\newcommand{\tL}{t_\mathcal{L}}
\newcommand{\tR}{t_\mathcal{R}}
\newcommand{\epsd}{\epsilon_\text{d}}
\newcommand{\MD}{M_\text{D}}
\newcommand{\tLu}{t_{\mathcal{L},\uparrow}}
\newcommand{\tRu}{t_{\mathcal{R},\uparrow}}
\newcommand{\tLd}{t_{\mathcal{L},\downarrow}}
\newcommand{\tRd}{t_{\mathcal{R},\downarrow}}
\newcommand{\To}{\mathcal{T}_\text{0}}
\newcommand{\As}{{\mathfrak A}}

\begin{document}
\title{Invariants of the single impurity Anderson model and
  implications for conductance functionals}

\author{F. Evers\inst{1,2,3} and P. Schmitteckert\inst{1,3} }
\shortauthor{F. Evers and P. Schmitteckert}
\institute{
\inst{1}{Institute of Nanotechnology, Karlsruhe Institute of
Technology (KIT), 76021 Karlsruhe, Germany}\\
\inst{2}{Institut f\"ur Theorie der Kondensierten Materie, 
KIT, D-76128 Karlsruhe, Germany}\\
\inst{3}{DFG-Center for Functional Nanostructures, 
KIT, D-76131 Karlsruhe, Germany}
}

\abstract{
An exact relation between the conductance maximum $G_0$  
at zero temperature and a ratio
of lead densities is derived within the framework of the single
impurity Anderson model: $G_0={\mathfrak R}[n] \frac{2e^2}{h}$, where 
${\mathfrak R}[n]=4\Delta N_{{\cal L},x}
\Delta N_{{\cal R},x}/(\Delta N_{{\cal L},x}+\Delta N_{{\cal
    R},x})^2$ and $\Delta N_{{\cal L},x}$,  $\Delta N_{{\cal R},x}$  
denote the excess density in the left/right lead at distance $x$ 
due to the presence of the impurity at the origin, $x=0$.  
The relation constitutes a parameter-free expression of the 
 conductance of the model in terms of 
the ground state density that generalizes an earlier result to the 
generic case of asymmetric lead couplings. It turns out that 
the specific density ratio, ${\mathfrak R}[n]$, 
is independent of the distance to the impurity $x$,
the (magnetic) band-structure and filling fraction of the contacting wires, 
the strength of the onsite interaction, the gate voltage and the
temperature.  Disorder induced backscattering in the contacting 
wires has an impact on ${\mathfrak R}$ that we discuss. 
Our result suggests that it should be possible, in principle,  
to determine experimentally the peak conductance of the Anderson impurity 
by performing a combination of measurements of ground-state densities. 
}

\date{\today}

\maketitle

\section{Introduction}
The single impurity Anderson model (SIAM) is 
one of the most important model system to understand 
correlation effects on electron transport through narrow constrictions 
and quantum dots. \cite{hewsonBook}
The reason is that it is a minimal model featuring 
the two most important interaction
induced phenomena in these systems, the Coulomb blockade 
and the Kondo effect. \cite{kondo64}
Both these effects manifest themselves in the local 
spectral function,  
$A_\text{d}(E)$,  
of the impurity (single level quantum dot)
as a triple peak structure, Hubbard side-bands and Abrikosov-Suhl
resonance in the centre. For this reason, the spectral function and derived
properties were in the focus of research for the last 40 years. 
 \cite{bulla08}
The charge susceptibility of the impurity 
has been obtained analytically already in the 1980ies
with Bethe-Ansatz methods. \cite{tsvelik83,andrei83}
It is not surprising that it received 
comparatively less attention than for instance 
the spin-susceptibility, simply because at the heart of the 
Kondo-effect is the screening of the impurity spin by conduction-band
electrons. The manifestation of the associated correlation 
effects in the {\it ground state density}, $n(x)$, 
is more subtle. Such correlations are experimentally  less
accessible, namely only as a as shift of the impurity 
induced Friedel oscillations. \cite{affleck08} 

The situation has changed recently, and the ground state 
density moved more into the active research focus. The reason is that the SIAM has 
become an important model system 
to investigate fundamental properties of the density functional
theory (DFT). \cite{mera10,evers11,troester12, bergfield12, liu12, stefanucci11,kurth13}
From the point of view of DFT, the SIAM is an ideal
test-bed, because of its analytical solvability and also because 
it allows for numerically highly accurate treatments based, e.g., 
on the numerical renormalization groug (NRG)\cite{bulla08} or the 
density matrix renormalization group (DMRG)\cite{white92,raas05,peters11}. 
In particular, the earlier Bethe-Ansatz results\cite{tsvelik83} 
could be used in order to 
invert the relation between the impurity occupation
$n(0)=N_\text{d}$ (summed over both spin directions) 
and its on site energy $\epsilon_\text{d}$
(``gate potential'') so as to obtain the exact 
exchange-correlation potential of this model 
analytically. \cite{bergfield12,liu12}

One important feature of the SIAM is that at zero temperature the 
Friedel sum rule holds true, which constitutes an exact relation 
between the ground-state density and the scattering phase 
of particles at the Fermi-energy\cite{hewsonBook}: $\delta_\text{F}=\pi N_\text{d}/2$.  
\footnote{
Quite generally, $\delta_\text{F}/\pi$ denotes the number of bound states 
(per spin) introduced by the impurity into the Fermi sea. 
In the SIAM, the impurity introduces a state when its on-site 
potential $\epsilon_\text{d}$ is reduced from infinity to a value below 
the Fermi-energy of the leads. In the wide band limit the occupation of 
this state ("extra bound charge") equals $N_\text{d}$. In the more general situation, 
$N_\text{d}$ gives a substantial part to the extra charge but additional 
contributions sitting on the lead sides neighboring the impurity will also exist. 
}
The relation is very
convenient, because most theoretical treatments focus on the
spectral-function, which is closely related to $N_\text{d}$, 
while experiments study the conductance, which is given 
via the identity (in units $2e^2/h$) 
\bea
G &=& \To \sin (\delta_\text{F})^2
\label{e1} \\
 \To&\equiv& \frac{4|t_{\cal L}|^2 |t_{\cal R}|^2}{(|t_{\cal L}|^2
  + |t_{\cal R}|^2)^2}. 
\label{e2}
\eea
Friedel's sum rule establishes the connection between the two. 
Here, $t_{\cal L,R}$ denote the hopping matrix
elements,  that connect the single site impurity with the left and
right leads. 
\footnote{Multiplication of nominator and denominator of Eq. (\ref{e2}) 
with the local density of states $\rho(E)$ on the contact site yields the familiar expression
\be
\To =  \frac{4\Gamma_{\cal L} \Gamma_{\cal R}}{\left(\Gamma_{\cal L}
  + \Gamma_{\cal R}\right)^2}
  \ee
since $\Gamma_{\cal L}=2\pi |t_{\cal L}|^2 \rho(E)$. 
}

In the case of symmetric coupling, 
$t_{\cal L}{=}t_{\cal R}{\equiv} t'$, $\To=1$ and  
 Eq. (\ref{e1})  has a remarkable property: it 
establishes a relation between conductance and density that is free of 
microscopic model parameters, e.g., the onsite interaction $U$ and
$t'$. 
Based on this observation an important conclusion was 
drawn\cite{troester12,bergfield12,stefanucci11}:  
For symmetric coupling, 
any Schr\"odinger-type effective single particle
theory, that produces  the correct ground state density 
also reproduces the conductance. 
The Kohn-Sham (KS) formulation of DFT is such an effective single
particle theory and therefore 
the frequently employed DFT-based transport scheme 
based on Landauer theory with KS-scattering states
\cite{arnold07,brandbyge02, cuevasBook},
is justified -- as long as the symmetric 
SIAM applies and exact ground state functionals can be used. 

In this work we provide an explicit formula for the
conductance, $G$, that expresses the prefactor $\To$, 
Eq. (\ref{e2}), in a parameter free way 
as a ground state density ratio in the general case
of asymmetric coupling, 
$t_{\cal L}\neq t_{\cal R}$. Specifically, 
we show that $\To$ is given by 
\bea
\To&=&\mathfrak{R}[n] \label{e3} \\
{\mathfrak R}[n]&=&\frac{4\Delta N_{{\cal L},x}
\Delta N_{{\cal R},x}}{(\Delta N_{{\cal L},x}+\Delta N_{{\cal
    R},x})^2}
\label{e4}
\eea
 with
$\Delta N_{{\cal L},x}, \Delta N_{{\cal R},x}$ 
denoting the excess density in the left/right lead at distance $x$ 
due to the presence of the impurity in the origin, $x=0$. 

Before we formally derive our result, we discuss several consequences.
1. Eq. (\ref{e3})  implies that $\mathfrak{R}$ is an invariant in the sense that it
does not depend on the distance of the measurement point $x$ to the 
impurity site. 
2.  The invariance statement is very general. It is valid for any wire length
(finite size or infinite) for any wire material (band-structure of a
single channel wire and band filling) 
and for arbitrary Hubbard interaction $U$. Moreover, it also holds in equilibrium at
non-zero temperature. The only condition is that the SIAM applies. 
3. Eq. (\ref{e3}) provides a pure density-functional (units $2e^2/h$)
\be
\label{e5}
{\cal G}[n]= \mathfrak{R}[n] \ \sin (\frac{\pi}{2} N_\text{d})^2
\ee
that constitutes  an explicit parameter-free expression of the zero-temperature 
conductance in terms of the ground state density. The relation
implies, in particular, the remarkable fact that the peak conductance can be
determined numerically and at least in principle also experimentally, 
by a proper combination of ground state charge density measurements. 
Therefore, the ratio $\mathfrak{R}$ introduced here 
enjoys a fundamental status similar, in a sense, 
to the familiar scattering phase $\delta_\text{F}$. 
4. Eq. (\ref{e5})  generalizes earlier statements for symmetric coupling --  
KS-based transport calculations employing the exact ground state
XC-functional reproduce the exact many body conductance -- 
to the experimentally very important case of asymmetric coupling.

\section{Model definition and analytical derivations}
The SIAM features  a single level quantum dot
\be
\label{e6}
 \HH_\text{QD} {=} \epsilon_\text{d} \hNd+ 
    U \left( \hat{n}_{\mathrm{d}\uparrow} - \frac{1}{2} \right) \left(
      \hat{n}_{\mathrm{d}\downarrow} 
- \frac{1}{2} \right), 
\ee
where $\hNd = \hat n_{\mathrm{d},\uparrow}+\hat n_{\mathrm{d},\downarrow}$
with $\hat n_{\mathrm{d},\sigma} =
\hat{d}_{\sigma}^{\dagger}
\hat{d}_{\sigma}^{\phantom{g}}$
and spin $\sigma=\uparrow,\downarrow$. 
The full model Hamiltonian is given by 
\begin{eqnarray}
\label{e7}
\HH &=& \HH_\text{QD} +\HH_\text{T}  +\sum_{\alpha=\mathcal{L,R}} \HH_{\alpha} 
\end{eqnarray}
where the coupling to leads with a length of $M$ sites is described via
\begin{eqnarray}
  \HH_{\alpha} &{=}& {-}\sum_{x,x'=1, \sigma}^{M-1}
 t_{x,x'} \left( c^\dagger_{\alpha\sigma,x}
  c_{\alpha\sigma,x'}^{\phantom{g}} {+} \text{h.c.}\right),\\
  \HH_\text{T} &{=}& - \sum_{\sigma,\alpha} t_\alpha \left( c_{\alpha\sigma,1}^{\dagger} d_{\sigma}^{\phantom{g}} + d_{\sigma}^{\dagger} c^{\phantom{g}}_{\alpha\sigma,1} \right) \mathrm{.} 
\end{eqnarray}
The operators 
$c^{\phantom{g}}_{\alpha\sigma,x}, c_{\alpha\sigma,x}^{\dagger}$ 
denote fermionic annihilation and 
creation operators at site $x$. 
The rotation that we will employ with respect to the lead degrees of
freedom, $\alpha={\cal L,R}$, does not mix spins. 
We subject the lead Hamiltonian (notation suppresses spin-index)
\bea
\sum_{\alpha}H_{\alpha} =-\sum_{x,x'=1}^{M-1}t_{x,x'} \sum_{\alpha}\
c^\dagger_{\alpha,x}c_{\alpha,x'} + \text{h.c.} 
\eea
to the standard rotation
\begin{eqnarray}
\label{e12}
\left( 
\begin{array}{c} 
c_{+,x} \\
c_{-,x} 
\end{array}
\right) 
&=&\frac{1}{\tilde t}
\left( 
\begin{array}{cc}
t_{\cal L} & t_{\cal R} \\
-t_{\cal R} & t_{\cal L}
\end{array}
\right)
\left( 
\begin{array}{c} 
c_{{\cal L},x} \\
c_{{\cal R},x} 
\end{array}
\right)
\end{eqnarray}
and analogously for 
$c^\dagger_{{\cal L},x}, c^\dagger_{{\cal R},x}$. 
The specific form (\ref{e12}) has been chosen such 
that the symmetric,``$+$''-channel couples to the quantum dot, 
with an effective coupling
$\tilde t = \sqrt{|t_{\cal L}|^2 + |t_{\cal R}|^2}$, 
while the anti-symmetric,``$-$''channel fully decouples. 
This decoupling naturally extends to multi-orbital impurities under the restriction
of proportional coupling.
After decoupling the model reads
\bea
\hat H &{=}& \hat H_\text{QD} + 
\sum_{\sigma} \tilde{t} ( c^\dagger_{+\sigma,1}d^{\phantom{g}}_\sigma {+}
d^\dagger_{\sigma}c^{\phantom{g}}_{+\sigma,1} ) + \hat H_{+}{+}\hat H_{-}\\
\hat H_{\beta} \!\!&{=}&\!\!\!\! \sum_{x,x'=1}^{M-1} \!\!t_{x,x'} \sum_{\sigma}
\left(  c^\dagger_{\beta\sigma, x}c^{\phantom{g}}_{\beta\sigma,x'} {+}
  \text{h.c.}\right), \beta = \pm.
\label{e13}
\eea

\section{Invariants} Since the ``$-$'' channel decouples, we have 
$[\hat H, \hat H_{-}]{=}0$, so the many-body eigenfunctions of 
$\hat H$ factorize into a non-interacting piece belonging to 
$\hat H_{-}$ and an interacting rest.  In addition, 
$[\hat H,\hat N_{-\sigma}]{=}[\hat H_{-},\hat N_{-\sigma}]{=}0$ implying
that the number of particles per spin in the ``${-}$''channel, 
$N_{-,\sigma}$, are
good quantum numbers that come in addition to the total particle
number, $N$.

\begin{figure}[t]
\includegraphics[width=0.48\textwidth]{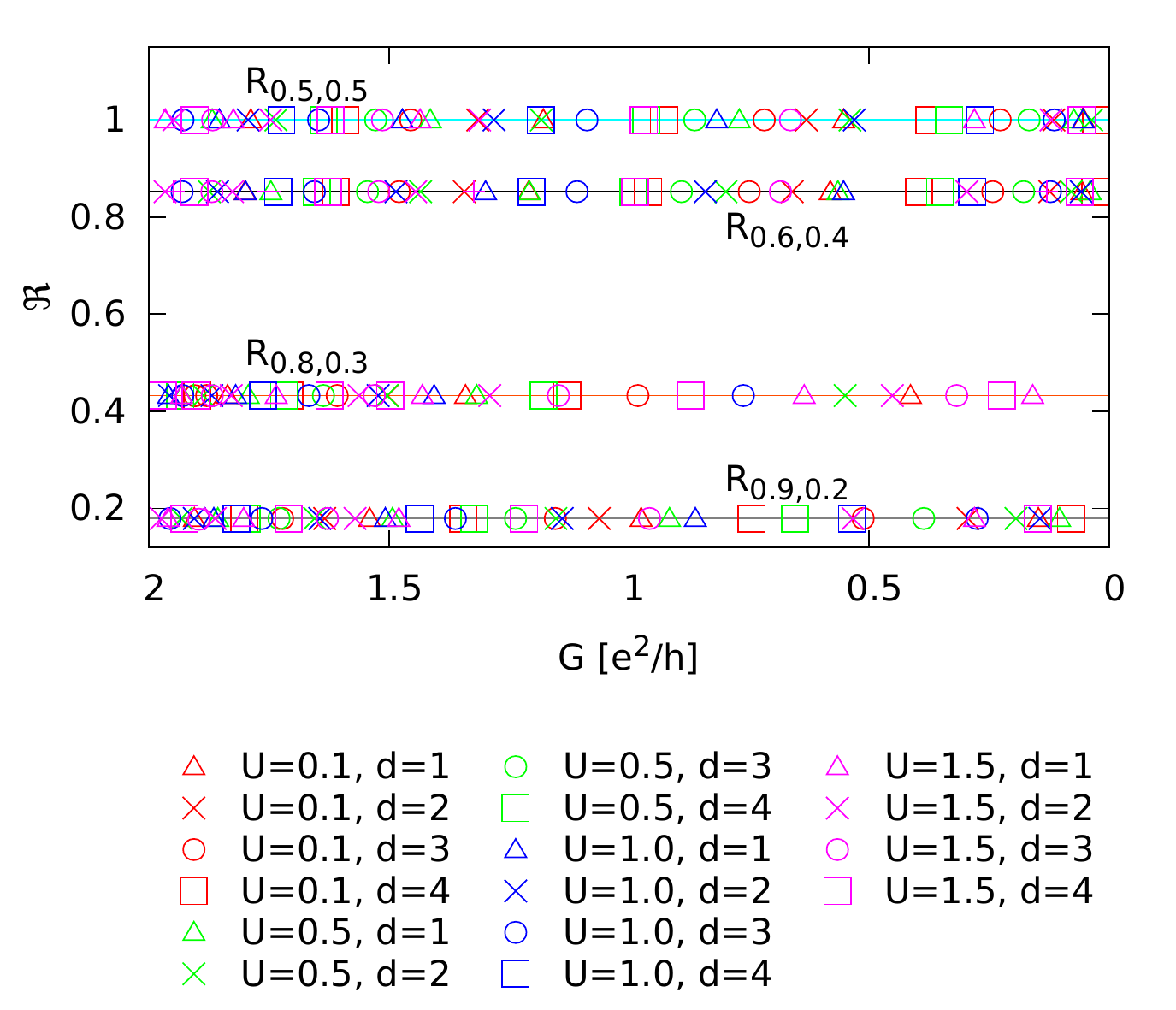}
\caption{ \ColorOnline \label{fig1} 
Asymmetry factor $\mathfrak{R}$ vs.\ the conductance 
for a single impurity coupled to $M{=}16$ left and right lead sites via 
($\tL=0.5$, $\tR=0.5$), ($\tL=0.6$, $\tR=0.4$), ($\tL=0.8$, $\tR=0.3$), and ($\tL=0.9$, $\tR=0.2$). 
$\mathfrak{R}$ was evaluated at distances $d=1, 2, 3, 4$ from the impurity site and with local interactions 
$U=\pm 1.5, \pm 1, \pm 0.5, 0.1$; the conductance was 
tuned by sweeping $\epsd$ between $0$ to 2.  
Data is obtained by evaluating \eqref{e4} (spin resolved) 
with DMRG ground-state densities at filling fraction 
corresponds to  32 fermions with total $S^z=0$.
Data points for $\mathfrak{R}$ show no visible deviations from $\To$
(solid lines) fully confirming the prediction Eq. (\ref{e3}). 
}
\end{figure}

We now investigate consequences of the existance of the two
invariants $N_{-\uparrow(\downarrow)}$. 
The thermodynamic excess densities at lead site $x$, 
that are induced by coupling to the impurity
at $x=0$, can be written as
\be
\Delta N_{\alpha,x} = \langle c^\dagger_{\alpha,x}c_{\alpha,x} \rangle
- \langle c^\dagger_{-,x}c_{-,x}\rangle, \quad \alpha={\cal L,R}.
\ee
The density  $\langle c^\dagger_{-,x}c_{-,x}\rangle$ of the
antisymmetric channel is our reference density. 
Note, that it is equivalent to the particle density in the leads
{\it before} wiring them to the impurity and therefore, 
can be determined directly 
in a suitable control measurement, at least in principle. 
The derivation of our result (\ref{e3}) proceeds by expressing
the orginal lead densities in terms of their rotated counterparts. 
The rotation is norm conserving, hence 
$
N_{{\cal L},x} {+} N_{{\cal R},x}  =
\langle c^\dagger_{+,x}c^{\phantom{g}}_{+,x}\rangle {+} \langle c^\dagger_{-,x}c^{\phantom{g}}_{-,x}\rangle
$
so that trivially 
\be
\Delta N_{{\cal L},x} + \Delta N_{{\cal R},x}  =
\langle c^\dagger_{+,x}c^{\phantom{g}}_{+,x}\rangle - \langle
c^\dagger_{-,x}c^{\phantom{g}}_{-,x}\rangle.
\label{e15}
\ee
Similarly, we derive for the difference 
$N_{\cal L}-N_{\cal R}{=}\langle c^\dagger_{{\cal L},x} c^{\phantom{g}}_{{\cal L},x} \rangle
 - \langle c^\dagger_{{\cal R},x} c^{\phantom{g}}_{{\cal R},x} \rangle$: 
\bea
\Delta N_{{\cal L},x} {-} \Delta N_{{\cal R},x} 
&=& {\As}  \left(\!\langle c^\dagger_{+,x}c^{\phantom{g}}_{+,x}\rangle{-}\langle
c^\dagger_{-,x}c^{\phantom{g}}_{-,x}\rangle\! \right)  \nonumber \\
&{-}& \To^{1/2} 
 \left(\! \langle c^\dagger_{+,x}c^{\phantom{g}}_{-,x}\rangle {+} \langle c^\dagger_{-,x} 
c^{\phantom{g}}_{+,x}\rangle \!\right)
\eea
where we have introduced the relative asymmetry 
 ${\As}=(|t_{\cal L}|^2{-}|t_{\cal R}|^2)/\tilde t^2$. 
Recalling Eq.  (\ref{e15}) we obtain the identity
\be
\label{e17}
\frac{\Delta N_{{\cal L},x} {-} \Delta N_{{\cal R},x} }
{\Delta N_{{\cal L},x} + \Delta N_{{\cal R},x}} 
{=} {\As}- 
\To^{1/2} \frac{\langle c^\dagger_{+,x}c^{\phantom{g}}_{-,x}\rangle {+} \langle c^\dagger_{-,x} 
c^{\phantom{g}}_{+,x}\rangle}{\Delta N_{{\cal L},x} + \Delta N_{{\cal R},x}}
\ee
for any  $x$ in the leads. The invariants $N_{-,\sigma}$ of the SIAM enter 
the final step of the proof. Since all eigenstates of $\hat H$
conserve the particle numbers in the ``${-}$''channel, the second term
on the rhs of Eq. (\ref{e17}) vanishes and we have 
\be
\label{e18}
\frac{\Delta N_{{\cal L}\sigma,x} {-} \Delta N_{{\cal R}\sigma,x} }
{\Delta N_{{\cal L}\sigma,x} + \Delta N_{{\cal R}\sigma,x}} 
= {{\As}}
\ee
with the spin-index restored. 
We arrive at the statement that the relative imbalance of the excess
densities (total or per spin)
equals everywhere the relative asymmetry of the junction irrespective
of the onsite interaction $U$, or temperature. 
A spin-resolved version of our results, Eqs. (\ref{e3},\ref{e4}),  
follows by squaring  each side of
Eq. (\ref{e18})  and then subtracting unity.  
Notice, that the derivation of Eq. (\ref{e18}) did not make 
specific reference to the band structure of the leads or 
the filling fraction. Hence, it holds for any 
lead band structure and in thermodynamic equilibrium at any temperature. 
Moreover, the derivation also 
does not make reference to the system size, $M$. Therefore, Eq. (\ref{e18}) 
is valid at any  length of the left/right hand side wires.

\subsection{Numerical check}
In Fig.~\ref{fig1} we compare the asymmetry ratio $\mathfrak{R}$ 
with $\To$ for different conductance values (obtained 
by varying $\epsd$ between zero and 2). 
We find $\mathfrak{R}$ (spin resolved), 
from the ground-state density that we obtain via a DMRG-calculation.
\footnote{ For our calculation we employed damped boundary 
conditions (DBC)\cite{bohr06} by scaling the hopping elements in
the leads by $\Lambda=0.8$ for the outermost $\MD=14$ bonds in  each lead.
The system was initialized using damping sweeps to ensure
convergence to the ground state in the presence of 
DBC\cite{bohr06}. We keep up to 3000 states per DMRG block.}
The data, Fig. \ref{fig1}, exhibits no visible deviations 
from Eq. (\ref{e3}) within the broad set of parameter values that 
was tested. The residual deviations exhibit an even/odd effect with respect 
to the distance from the impurity; they are  
smaller than $10^{-5}$ (odd) and  $10^{-8}$ (even) and can be 
attributed to numerical uncertainties in the DMRG-density.  

\begin{figure}[t]
\includegraphics[width=0.45\textwidth]{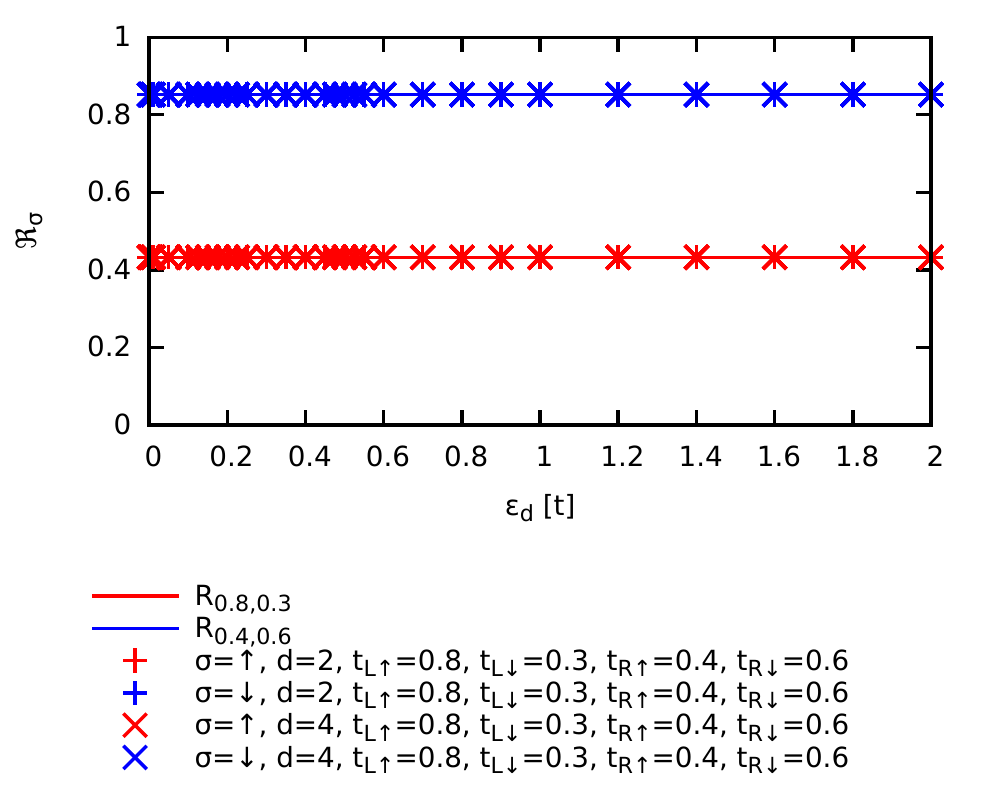}
\caption{\ColorOnline \label{fig2} 
Asymmetry factor $\mathfrak{R}_\sigma$ vs.\ the local potential $\epsd$ 
for a single impurity coupled to ${M}{=}16$ left and right lead sites,
via spin dependent hybridization
$\tLu=0.8$, $\tLd=0.3$  and $\tRu=0.4$, $\tRd=0.6$
measured at distances $d=2, 4$ from the impurity site and for local interactions of
$U=0, 0.01. 0.1, 0.3, 0.5, 0.7, 1.0, 1.5 2.0, 3.0, 5.0$ for each values of $\epsd$,
and 32 fermions with total $S^z=0$. 
}
\end{figure}

\section{Extension to magnetic leads}
 In recent measurements of the magneto-resistance of individual
molecules a spin-dependent coupling $t_{\alpha,\sigma}$ was introduced
to explain the experiments and motivated via DFT-based transport
studies. \cite{schmaus11,bagrets12}. In this spirit, we  generalize our
model now including the possibility of magnetic electrodes:  
\begin{eqnarray}
  \HH_{\alpha} &{=}& {-}\sum_{x,x'=1, \sigma}^{M-1}
 t_{x,x',\sigma} \left( \hat{c}_{\alpha\sigma,x}^{\dagger}
  \hat{c}_{\alpha\sigma,x'}^{\phantom{g}} {+} \text{h.c.}\right)\\
  \HH_\text{T} &{=}& - \sum_{\alpha,\sigma} t_{\alpha\sigma} \left(
    \hat{c}_{\alpha\sigma,1}^{\dagger} \hat{d}_{\sigma}^{\phantom{g}} 
+ \hat{d}_{\sigma}^{\dagger} \hat{c}_{\alpha\sigma,1}^{\phantom{g}} \right) \mathrm{.} 
\end{eqnarray}
The rotation \eqref{e12}  operates on each spin sector separately, and
each spin direction has its own invariant, $N_{-\sigma}$. Hence, 
we immediately conclude
${{\To}_\sigma}=\mathfrak{R}_{\sigma}[n_\sigma] $
($n_\sigma$: particle number density 
per spin in the ground-state), 
\bea
{\mathfrak R}_\sigma[n_\sigma]&=&\frac{4\Delta N_{{\cal L}\sigma,x}
\Delta N_{{\cal R}\sigma,-x}}{(\Delta N_{{\cal L}\sigma,x}+\Delta N_{{\cal
    R}\sigma,-x})^2}
\label{e21} \\
 {\To}_\sigma &\equiv& \frac{4|t_{{\cal L}\sigma}|^2 |t_{{\cal R}\sigma}|^2}{(|t_{{\cal L}\sigma}|^2
  + |t_{{\cal R}\sigma}|^2)^2}. 
\label{e22}
\eea

Again, we subject our analytical findings to a numerical test. 
In  Fig.~\ref{fig2} we display asymmetry ratios $\mathfrak{R}_\sigma$  
for asymmetric, spin-dependent 
couplings to non-magnetic wires. The data fully  confirms our analysis,
Eqs. (\ref{e21},\ref{e22}).

\section{Discussion}
The results presented so far crucially rely upon the existance of two 
extra invariants in the SIAM, $N_{-\uparrow (\downarrow)}$. They originate
from two special features: (i) there is only a single level  
(more precisely: proportional coupling is required in the sense of Ref.\cite{meir92}) 
(ii) all leads have an identical electronic structure. 
Since both of these features capture important aspects but not all 
of physical reality, one may ask about the effect of perturbations. 

\subsection{Competing orbitals} A general discussion of possible 
multi-orbital terms that could complement $\hat H_\text{QD}$ in 
realistic situations is extremely complicated and not indicated here. 
Instead, we can recall that the density modulation in the leads, $\Delta N_{\cal L,R}$,  
is due to the change in occupation 
of the frontier-levels of the QD and  as such a Fermi-surface effect. 
Therefore, in the spirit of the Fermi-liquid theory any microscopic description 
is justified that correctly represents this low-energy sector. 
And because of this, the SIAM is indeed very successful when describing the ($S{=}1/2$) 
Kondo-effect as it is frequently observed in experiments\cite{scott10}; 
this includes details like the temperature scaling. 

\subsection{Disorder in the leads}
Here, we offer a first test of the assumption (ii) and ask about the effect of small differences 
in the band structures of the left/right  wires. We subject the leads to onsite disorder, 
$\hat H_\text{dis}$, where the local energies are drawn from a box
distribution of width $W$.
The term in $\hat H$ that introduces the difference between left/right leads reads 
\be
\hat H_\text{dis}{=}\!\!\sum_{\sigma, x=1}^{M}v_{x} 
\left( 
c^\dagger_{{\cal L}\sigma,x}c^{\phantom{g}}_{{\cal L}\sigma,x} {-} c^\dagger_{{\cal
    R}\sigma,x} c^{\phantom{g}}_{{\cal R}\sigma,x} \right)
{=} \hat H_\text{as} {+} \hat V_\text{T} 
\ee
where 
\bea
\hat H_\text{as} &=&  {\As} \sum_{\sigma, x} v_x \left(
  c^\dagger_{+\sigma,x}c^{\phantom{g}}_{+\sigma,x} {-}
c^\dagger_{-\sigma,x}c^{\phantom{g}}_{-\sigma,x} \right)   \nonumber \\
\hat V_\text{T} &=&-\To^{1/2} \sum_{\sigma,x} v_x
\left(  c^\dagger_{+\sigma,x}c^{\phantom{g}}_{-\sigma,x} +  c^\dagger_{-\sigma,x} 
c^{\phantom{g}}_{+\sigma,x} \right).
\eea
The first term, $\hat H_\text{as}$, still respects the basic symmetries of the SIAM outlined
below Eq. (\ref{e13}).  Its most important effect is to modulate 
$\langle c^\dagger_{\beta\sigma,x}c^{\phantom{g}}_{\beta\sigma,x}\rangle$ 
in space with an amplitude that vanishes for symmetric coupling, 
when ${\As}\to 0$.  
\begin{figure}[tbh]
\includegraphics[width=0.45\textwidth]{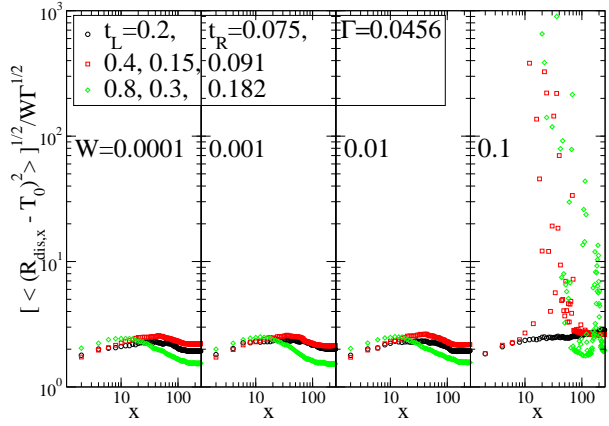}
\caption{\ColorOnline \label{fig3} 
Mean deviation of ${\mathfrak R}_{\text{dis},x}$ (see text) 
from $\To$ normalized by the disorder strength $W$ 
and the level broadening $\Gamma^{1/2}$. 
Data is given at even sites away from the impurity  
for increasing disorder $W=10^{-4},10^{-3}, 10^{-2}, 0.1$ and 
different couplings $t_{\cal L,R}$. ($\To=0.43235$ is fixed, $U=0$, $M{=}256$
and  $\epsilon_\text{d}=0.1$.)  
It is seen here that even in the presence of moderate disorder
${\mathfrak R}_{\text{dis},x}$ can give an
estimate for $\To$ with accuracy $\sim W\Gamma^{1/2}$ if one
focusses on the near impurity region, where $x$ is much 
smaller than the mean free path, $\ell$. 
(We recall that in the present lead model 
$\ell \approx 100/W^2$.\cite{kramer93}) 
}
\end{figure}

The disorder term $\hat V_\text{T}$ includes tunnelling
between the ``$+$'' and ``$-$'' channels, so that 
$N_{-\uparrow(\downarrow)}$ are no longer conserved in its presence; 
the second term on the rhs of Eq. (\ref{e17}) no longer vanishes, in general. 

To complement the qualitative discussion, we have performed
a numerical simulation in order to quantify the effect of the 
symmetry-breaking terms.  We define an observable 
${\mathfrak R}_{\text{dis},x}$ which is similar to ${\mathfrak R}_{x}$ of 
 Eq. (\ref{e3}), with the difference that we approximate 
$\langle c^\dagger_{-,x}c^{\phantom{g}}_{-,x}\rangle
\approx (\bar N_{{\cal L},x}{+}\bar N_{{\cal R},x})/2$ where 
$\bar N_{{\cal L},x}, \bar N_{{\cal R},x}$ denote 
 the particle density in the the leads in the absense of a coupling to the
impurity site. The object ${\mathfrak R}_{\text{dis},x}$ is
interesting to study since it is an explicit density functional 
that reduces to the exact expression in the
limit $W{\to} 0$ and is directly accessible to experimental
measurements, at least in principle. 
Importantly, our simulations, Fig. \ref{fig3}, suggests 
that in the case of asymmetric coupling, 
${\mathfrak R}_\text{dis}$ exhibits fluctuations about $\To$ 
of a size  $\sim c W\sqrt{\Gamma}/|t_{{\cal L},{\cal R}}|^{3/2}$ with a factor $c\approx 2$. 
This result is encouraging, because it shows that disorder effects 
can be well controlled by going to weakly coupled QDs. 
Hence, we believe that there are promising prospects 
that even in the presence of moderate disorder one can still 
estimate the maximum conductance with good accuracy in experiments 
by performing a sequence of ground-state  density measurements.


\acknowledgements
{\it Acknowledgements:} FE would like to express his gratitude to the IAS at the
HUJI and its staff for their warm hospitality while this work was
performed. Also, FE thanks J. C. Cuevas, N. Andrei, A. Schiller,
O. Entin-Wohlman,  E. Rabani, U. Peskin and A. Aharony for inspiring discussions. 
Most of calculations were performed on the compute cluster of the YIG group of Peter Orth.



\begin{thebibliography}{10}
\expandafter\ifx\csname url\endcsname\relax\def\url#1{\texttt{#1}}\fi

\bibitem{hewsonBook}
\Name{Hewson A.~C.} \Book{The Kondo Problem to Heavy Fermions} (Cambridge
  Studies in Magnetism, New York, N.Y.) 1995.
  
\bibitem{kondo64}
\Name{Kondo J.} \REVIEW{Progress of Theoretical Physics}{32}{1964}{37}.

\bibitem{bulla08}
\Name{Bulla R., Costi T. \and Pruschke T.} \REVIEW{Rev. Mod.
  Phys.}{80}{2008}{395}.

\bibitem{tsvelik83}
\Name{Tsvelik A. \and Wiegmann P.} \REVIEW{Adv. in Phys.}{32}{1983}{453}.

\bibitem{andrei83}
\Name{Andrei N., Furuya K. \and Lowenstein J.~H.} \REVIEW{Rev. Mod.
  Phys.}{55}{1983}{331}.

\bibitem{affleck08}
\Name{Affleck I., Borda I. \and Saleur H.} \REVIEW{Phys. Rev.
  B}{77}{2008}{180404(R)}.

\bibitem{mera10}
\Name{Mera H. \and Niquet Y.} \REVIEW{Phys. Rev. Lett.}{105}{2010}{216408}.

\bibitem{evers11}
\Name{Evers F. \and Schmitteckert P.} \REVIEW{Phys. Chem. Chem.
  Phys.}{13}{2011}{14417}.

\bibitem{troester12}
\Name{Tr{\"o}ster P., Schmitteckert P. \and Evers F.} \REVIEW{Phys. Rev.
  B}{85}{2012}{115409}.

\bibitem{bergfield12}
\Name{Bergfield J.~P., Liu Z., Burke K. \and Stafford C.~A.} \REVIEW{Phys. Rev.
  Lett.}{108}{2012}{066801}.

\bibitem{liu12}
\Name{Liu Z., Bergfield J.~P., Burke K. \and Stafford C.~A.} \REVIEW{Phys. Rev.
  B}{85}{2012}{155117}.

\bibitem{stefanucci11}
\Name{Stefanucci G. \and Kurth S.} \REVIEW{Phys. Rev.
  Lett.}{107}{2011}{216401}.

\bibitem{kurth13}
\Name{Kurth S. \and Stefanucci G.}
\REVIEW{Phys. Rev. Lett.}{111}{2013}{ 030601}

\bibitem{white92}
\Name{White S.~R.} \REVIEW{Phys. Rev. Lett.}{69}{1992}{2863}.

\bibitem{raas09}
\Name{Raas G. \and Uhrig G.}\REVIEW{Euro. Phys. Journal B}{45}{2005}{293}.

\bibitem{peters11}
\Name{Peters R.} \REVIEW{Phys. Rev. B}{84}{2011}{075139}.

\bibitem{arnold07}
\Name{Arnold A., Weigend F. \and Evers F.} \REVIEW{J. Chem.
  Phys.}{126}{2007}{174101}.

\bibitem{brandbyge02}
\Name{Brandbyge M., Mozos J.-L., Ordejon P., Taylor J. \and Stokbro K.}
  \REVIEW{Phys. Rev. B}{65}{2002}{165401}.

\bibitem{cuevasBook}
\Name{Cuevas J.~C. \and Scheer E.} \Book{Molecular Electronics: An Introduction
  to Theory and Experiment} World Scientific Series in Nanotechnology and
  Nanoscience (World Scientific) 2010.

\bibitem{bohr06}
\Name{Bohr D., Schmitteckert P. \and W\"olfle P.} \REVIEW{Europhys.
  Lett.}{73}{2006}{246}.

\bibitem{schmaus11}
\Name{Schmaus S., Bagrets A., Nahas Y., Yamada T.~K., Bork A., Bowen M.,
  Beaurepaire E., Evers F. \and Wulfhekel W.} \REVIEW{Nature
  Nanotechnology}{6}{2011}{185}.

\bibitem{bagrets12}
\Name{Bagrets A., Schmaus S., Jaafar A., D. K., Kazu T., Alouani M., Wulfhekel
  W. \and Evers F.} \REVIEW{Nano Lett.}{12}{2012}{5131}.

\bibitem{meir92}
\Name{Meir Y. \and Wingreen N.} \REVIEW{Phys. Rev. Lett.}{68}{1992}{2512}.

\bibitem{scott10}
\Name{Scott G. \and Natelson D.} \REVIEW{ACS Nano}{4}{2010}{3560}.

\bibitem{kramer93}
\Name{Kramer B. \and MacKinnon A.} \REVIEW{Rep. Prog. Phys.}{56}{1993}{1469}.

\end{thebibliography}
\end{document}